\documentclass[preprint,12pt]{elsarticle}
\usepackage{graphicx}
\usepackage[linesnumbered,boxed,ruled]{algorithm2e}

\journal{Simulation Modelling Practice and Theory}

\begin{document}

\begin{frontmatter}

\title{Resource Availability-Aware Advance Reservation for Parallel Jobs with Deadlines}

\author[ynu]{Bo Li\corref{cor1}}
\ead{libo@ynu.edu.cn}
\author[ynu]{Yijian Pei}
\author[whut]{Bin Shen}
\author[ynu]{Hao Wu}
\author[ynu]{Min He}
\author[ynu]{Jundong Yang}

\cortext[cor1]{Corresponding author}
\address[ynu]{School of Information Science and Engineering, Yunnan University, Kunming 650091, China}
\address[whut]{School of Electrical and Information Engineering, Wuhan Institute of Technology, Wuhan 430073, China}

\begin{abstract}
Advance reservation is important to guarantee the quality of services of jobs by allowing exclusive access to resources over a defined time interval on resources. It is a challenge for the scheduler to organize available resources efficiently and to allocate them for parallel AR jobs with deadline constraint appropriately. This paper provides a slot-based data structure to organize available resources of multiprocessor systems in a way that enables efficient search and update operations, and formulates a suite of scheduling policies to allocate resources for dynamically arriving AR requests. The performance of the scheduling algorithms were investigated by simulations with different job sizes and durations, system loads and scheduling flexibilities. Simulation results show that job sizes and durations, system load and the flexibility of scheduling will impact the performance metrics of all the scheduling algorithms, and the $PE\; Worst Fit$ algorithm becomes the best algorithm for the scheduler with the highest acceptance rate of AR requests, and the jobs with the $First Fit$ algorithm experience the lowest average slowdown. The data structure and scheduling policies can be used to organize and allocate resources for parallel AR jobs with deadline constraint in large-scale computing systems.
\end{abstract}


\end{frontmatter}


\section{Introduction}
\label{sec:introduce}

Grid-like massive Internet computing platforms have emerged as an essential infrastructure for scientific and commercial applications and made it possible for "flexible, secure, coordinated resource sharing among dynamic collections of individuals, institutions, and resources"[1]. In order to guarantee the QoS of Grid applications in such environments, advance reservation(AR) is incorporated into many Grid systems, such as GARA\cite{foster2004end}, Nimrod/G\cite{buyya2000nimrod} and G-QoSM\cite{al2002g}, and many mainstream parallel scheduler, such as Maui\cite{jackson2001core}, load sharing facility(LSF)\cite{lsf} and protable batch system(PBS)\cite{bode2000portable}. Advance reservation makes the execution of grid jobs more predictable by reserving a particular resource capability over a defined time interval on local resources, and has been widely used for many kinds of jobs, such as single processor jobs\cite{castillo2011online}, parallel jobs\cite{nurmi2009probabilistic},
mixed-parallel applications\cite{aida2009scheduling}, co-allocation jobs\cite{castillo2009resource},
bag-of-tasks\cite{lee2010rescheduling}, and workflow\cite{cucinotta2009advance}. Advance reservation has been an important area of interest in the Grid community.

In order to reserve resources in such AR systems, a user must submit a request to the system by specifying a set of parameters such as number of processing elements needed,ready time, duration and deadline. For such an AR request, the job cannot start until its ready time and it must be completed by its deadline. Upon receiving such an AR request, it's the task of the scheduler to decide if there are sufficient available resources such that the request can be completely executed within the interval of its ready time and deadline. Considering that there are so many resources and optional allocations to check in a large-scale computing system, it's quite a challenge for the scheduler to organize available resources efficiently and to allocate resources for dynamically arriving AR requests appropriately: the scheduling procedure itself will impact the ability and efficiency of the scheduler to manage resources and to schedule a great number of jobs with various requirements, and the scheduling decision will also impact the performance perceived by users and service providers. For users, the fraction of AR requests accepted by the scheduler and the turnaround time are important measures of how well their service requests are treated\cite{li2007performance,snell2000performance}. For clusters, the acceptance of new reservations will fragment a continuous range of resource into pieces, and thus reduce the potential scheduling opportunities and results in lower utilization\cite{naiksatam2007elastic,margo2008impact}. The key challenges here lies in two aspects: (1) to develop efficient data structure to organize available resources for AR requests in a way that enables efficient search and update operations; (2) to develop a group of scheduling algorithms or policies that improve the performance perceived by users and providers.

In the literature, many data structures(such as array\cite{burchard2005analysis}, linked-list\cite{xiong2005linked},
trees\cite{wang2002bandwidth,burchard2005analysis,nie2010flexible} and queues\cite{brown1988calendar,Sulistio2009}) and scheduling algorithms(\cite{netto2010sla,naiksatam2007elastic,balakrishnan2010sla,nie2010flexible}) have been proposed for advance reservations and widely studied. However, these data structures and scheduling algorithms are only suitable for single- or multi-processor AR jobs with immediate deadline constraint. For such kind of AR jobs, they must be scheduled to run at their ready time and their deadlines are immediate(i.e., deadline=ready time+duration). However, if the AR requests are not strictly asked to begin to run at their ready times, they can begin to run at any time within its ready time to its latest start time(i.e., =deadline-duration). Such kind of AR request is more general than those with immediate deadlines, and makes it more flexible and complicated for the scheduler to organize available resources, to control admission and to schedule. As a result, all those existing data structures and scheduling algorithms for AR requests with immediate deadlines are not suitable for these with general deadlines.

Up to now, only few researches have been done for AR jobs with general deadlines. In \cite{xu2004efficient} and \cite{castillo2011online,castillo2007design}, the authors investigated the problem of how to allocate a single-channel(or single-processor) AR job with general deadline constraint to $n$ single-processor servers. In those works, because each job only needs to be reserved on a single-processor server, it is not necessary to allocate more than one idle intervals across multiple servers for them simultaneously, thus all the algorithms in those works only considered the temporal constraint, without considering the scheduling of AR jobs with more than one resources simultaneously.

Despite the fact that existing data structures and algorithms have been widely used for AR requests, they are not suitable for parallel AR jobs with general deadlines in large-scale multiprocessor systems. In this paper, we investigated the problem of how to manage and allocate multiprocessor resources for parallel AR jobs with general deadline constraint. Different from existing data structures and scheduling policies designed for scheduling single-processor deadline-constraint AR jobs to $n$ single processor servers, or for scheduling parallel AR jobs with immediate deadlines to a multiprocessor system, in this work we proposed a new data structure and scheduling policies to organize the availability of resources in a large-scale multiprocessor system and to allocate them appropriately for parallel AR jobs with general deadline constraint. The main contribution of this work include:
\begin{itemize}
  \item {Proposed a new data structure to organize available resources efficiently in multiprocessor systems for \emph{single- or multiple-processor} AR requests with \emph{immediate or general }deadline constrains;}
  \item{Proposed a set of operations for the data structure to enable efficient search and update operations;}
\item{Proposed a set of scheduling policies for the data structure to allocate resources for AR requests, and investigated their performance via simulation. New scheduling policies can be added into the data structure flexibly.}
\end{itemize}

The rest of this paper is organized as follows. We discuss related work in Section 2 and describe the model for scheduling parallel AR jobs with general deadlines in a multiprocessor system in Section 3. In Section 4 we introduce a slot-based data structure to organize the availability of resources in a way that enables efficient adding, deleting and searching operations. In Section 5 we provide a suite of scheduling algorithms for parallel AR requests with general deadline constraint and present a comprehensive performance evaluation study of the algorithms by simulations, and we conclude the paper in Section 6.

\section{Related work}
\label{sec:introduce}

Many data structures and scheduling algorithms have been proposed for advance reservations. Most of them are suitable for AR requests with immediate deadlines, and only few of them were specifically designed for AR requests with general deadlines. For AR request with immediate deadlines, such data structure as array\cite{burchard2005analysis}, linked-list\cite{xiong2005linked},
trees\cite{wang2002bandwidth,burchard2005analysis,nie2010flexible} and queues\cite{brown1988calendar,Sulistio2009} have already been widely studied. These data structures are primarily used for admission control and focused on finding out whether it's feasible for the scheduler to accept an AR request to start at a definite time and keep on running for a given period. In \cite{Sulistio2009} the author presents a good summary and comparison of them when they are used for single- or multi-processor AR jobs with immediate deadline constraint. However, they are not specifically designed for AR requests with general deadline constraint.
Based on existing scheduling theory and algorithms for jobs with or without deadlines, some variants of scheduling algorithms for jobs with advance reservations have been studied in Grid-like systems\cite{netto2010sla,naiksatam2007elastic,balakrishnan2010sla,nie2010flexible} and their impact on the users and the systems were investigated in terms of turnaround time, slowdown, or utilization.

Different from existing plentiful researches for AR requests with immediate deadlines, only few works have been done for AR jobs with general deadlines. In \cite{xu2004efficient},the problem of how to reserve optical bursts on wavelength channels whose bandwidth may become fragmented with idle intervals was proposed. By using concepts from computational geometry,the author maps each idle interval and each burst as a point on a two-dimensional plane,then the points for idle intervals were organized into a search tree and several algorithms,such as
Min-SV, Max-SV, Min-Ev, Max-EV and Best-fit,were proposed for reserving bursts with and without fiber delay lines. Based on the concept and algorithms in \cite{xu2004efficient}, in \cite{castillo2011online,castillo2007design}
the author adapted them for scheduling \emph{single-processor} AR jobs with general deadline constraint to $n$ \emph{single-processor servers}. In those works, because each job only needs to be reserved on a single-processor server, it is not necessary to allocate more than one idle intervals across multiple servers for them simultaneously, thus all the algorithms in those works only considered the temporal constraint, without considering the scheduling of AR jobs with more than one resources simultaneously. Moreover, for different scheduling policies in \cite{castillo2011online,castillo2007design}, the data structure used for storing the availability information of the resources and the method for finding out the appropriate interval are different. This limits the flexibility of the data structures to support new scheduling policies. In contrast, the data structure proposed in this paper can support different scheduling policies flexibly, without changing the data structure itself and the method of finding appropriate resources for the requests.

\section{Problem description}
\label{sec:model}
\begin{figure}[htp]
\centering
\includegraphics[width=0.9\textwidth]{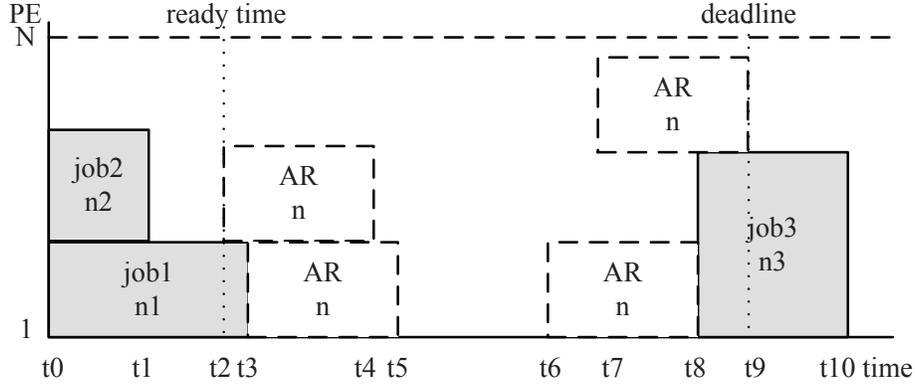}
\caption{An example of reserving processing elements for  a new advance reservation request with general deadline constraint. Assume the new AR request arrives at $t0$, when there are 2 running jobs(job1 and job2) and one reserved job(job3), and the AR request needs $n$ processing elements and should be processed within its ready time($t2$) and its deadline($t9$). Four feasible allocation for the AR request are illustrated.}
\label{fig:fig1}
\end{figure}

The computing environment is a parallel system, e.g., clusters or massively parallel processing machines, consisting of a group of space-shared processing elements $\{PE_1, PE_2, ..., PE_n\}$, with the total number of $n$. For simplicity, we assume the PEs are homogeneous. Each machine has a local resource management system capable of supporting advance reservations for local or external jobs.
Figure \ref{fig:fig1} shows an example schedule of a parallel AR job with general deadline constraint. Assume the request of the AR job arrives at $t_0$. The request asks the scheduler to allocate PEs during the ready time and the deadline so as the job can run for its duration. On receiving this request, the scheduler will evaluate  whether they are enough resources available for the job so as to meet its deadline. If so, the scheduler will allocate and reserve them for the job; otherwise,the request will be declined. Moreover, if there are more than one allocations that can satisfy the request at the same time, only one of them will be chosen based on some criteria or policies.

In this paper, each AR request with deadline is characterized by a five-parameter tuple $(t_a, t_r, t_{du}, t_{dl}, n_{pe})$, where:
\begin{enumerate}
    \item $t_a$ is the arrival time of the request;
    \item $t_r (\ge t_a)$ is the ready time, i.e., the earliest start time of the job. When $t_r>t_a$ is permitted, advance reservations are supported by the scheduler; Otherwise, only immediate reservations are permitted;
  \item $t_{du}$ is the duration of the job, i.e., the amount of time needed by the job when running on current cluster;
  \item $t_{dl} (\ge t_r+t_{du})$ is the deadline, i.e., the latest time by which current job must be completed. If $t_{dl}=t_r+t_{du}$, the deadline is immediate and we refer to this problem as scheduling with immediate deadline. If $t_{dl}\ge t_r+t_{du}$, the deadline is general and we refer to this problem as scheduling with general deadline; and
  \item $n_{pe}$ is the number of PEs required by the request.
\end{enumerate}

In Figure \ref{fig:fig1}, at $t_0$, there are respectively two running jobs(job1 and job2) and one reserved job(job3) on the cluster. The scheduler can try to allocate the job to start at any time from the ready time (i.e., $t2$) to the latest start time, i.e., $t7 (=t_{dl}-t_{du})$, and then check whether there are enough PEs for the job to begin to run at the selected start time for $t_{du}$.

In this paper, we assume all AR jobs arrive dynamically and they are non-preemptive and non-malleable, i.e., they must run till completion once they start execution and their requirements on resources, such as the number of PEs, can not be changed. Compared with preemptive and malleable AR jobs\cite{naiksatam2007elastic}, such kind of non-preemptive and non-malleable AR jobs are more difficult to tackle for the scheduler. It appears to be NP-complete to schedule them under deadline constraint even for very restricted cases, and there are not optimal online scheduling algorithms for them\cite{pinedo2008scheduling}. In order to schedule this kind of AR jobs, heuristics are left for the scheduler.

In the scheduling, both of the procedure to decide and the decision itself made by the scheduler are important. Because there are so many PEs and optional start times to check, the scheduling procedure itself will impact the ability and efficiency of the scheduler to manage a large-scale distributed resources and to schedule a great number of jobs with various requirements, and the scheduling decision will also impact the performance perceived by users and service providers. In the following sections, an efficient data structure and operations, and a suite of scheduling strategies will be proposed to manage and allocate resources for AR requests with deadline.

\section{Data Structure and operations}

Similar to the variable slot data structure in \cite{kunrath2008towards}, for each cluster, we represent the resources allocated for each running or reserved job as an rectangle and record the availability of a cluster as a set of $\{time, PEs\}$ pairs, where $time$ means at which the state(i.e., busy or idle) of the PEs change, and $PEs$ means the identities of the PEs who are busy at $time$. If $PEs$ is null($\o$), it means at that time all busy PEs recorded in the previous time slot are set free.

In order to record and manage the identity of PEs occupied in every time period, a linked list-based data structure \emph{AvailRectList} was proposed. When a new job is allocated with it start time($t_s$), it end time($t_e$) and its PEs($PE_{job}$), the records within the interval $[t_s,t_e)$ will be updated by adding $PE_{job}$ to their PEs, inferring that $PE_{job}$ will be used by the job from $t_s$ to $t_e$. Accordingly, when the job is completed, $PE_{job}$ will be released and subtracted from the records within $t_s$ to $t_e$. In this way, at any time, the PEs already occupied for running or reserved jobs are known, and we can check for the availability of PEs in any given time interval.

Additionally, to simplify the process of querying the time slots at which the states of PEs change and the availability of PEs, an auxiliary sorted set-based structure, {\em TimeSet}, was used. As the records in \emph{AvailRectList} change, {\em TimeSet} will be updated synchronously.

In order to support advance reservations with deadline, the data structure needs to perform three basic operations: adding an allocation, deleting an allocation and searching for available allocations for AR requests.

\subsection{Adding and deleting an allocation}

Before adding an allocation, we assume a search operation(see Subsection \ref{sec:search}) has already been done and the start time($t_s$), the end time($t_e$) and the PEs of the job($PE_{job}$) have already been allocated. The adding operation is described in Algorithm $addAllocation(t_s, t_e, PE_{job})$. The core of this operation is to update the records in the interval $[t_s,t_e)$ by adding $PE_{job}$ to their PEs. If \emph{AvailRectList} is empty or the earliest time of the records is greater than $t_e$, just need to add $\{t_s, PE_{job}\}$ and $\{t_e, \o\}$ into \emph{AvailRectList};Otherwise,we should find all records in the interval $[t_s,t_e)$ and update their PEs by adding them with $PE_{job}$(line \ref{alg:line:p3}-\ref{alg:line:p4}). After updating, it is possible that the PEs of the record of $t_s$ or $t_e$ become the same as that of the record of the time slot just before, or that $t_s$ is the earliest time slot and the PEs of the record of $t_s$ are null. In such cases, the records of $t_s$ and/or $t_e$ are redundant and should be cleaned((line \ref{alg1:line:p7}). When a job finishes, a deleting operation $deleteAllocation(t_s, t_e, PE_{job})$ is called immediately. It applies to the same principle as adding a new one but to update the records in the interval $[t_s, t_e)$ by subtracting $PE_{job}$ from their PEs.

\begin{algorithm}[htbp]
\caption{addAllocation($t_{s}$,$t_{e}$,$PE_{job}$)}
\label{alg:add}

\eIf{$AvailRectList$ is empty \em \emph{OR} $TimeSet.first>t_e$}
    {
        $AvailRectList.addall$(\{$t_{s}$,$PE_{job}$\},\{$t_{e}$,\o\})\;
        \label{alg:line:p1}
    }
    {
        find all records within $[t_s,t_e)$ in $AvailRectList$\;\label{alg:line:p3}
        update the PEs of the records found by adding them with $PE_{job}$\;\label{alg:line:p4}
    \label{alg1:line:p6}
    }
clean possible redundant records\;
\label{alg1:line:p7}
\end{algorithm}

\begin{algorithm}[htbp]
\caption{deleteAllocation($t_{s}$,$t_{e}$,$PE_{job}$)}
\label{alg:delete}
find all records within $[t_s,t_e)$ in $AvailRectList$\;
update the PEs of the records found by subtracting them with $PE_{job}$\;
clean possible redundant records\;
\end{algorithm}

The complexity of the \emph{addAllocation()} or the \emph{deleteAllocation()} operation is analyzed as follows. Suppose there are $n$ records in $AvailRectList$. For the \emph{addAllocation()} or the \emph{deleteAllocation()} operation, we need to update the records within $[t_s,t_e)$. Assume there are $n'$ records within
$[t_s,t_e)$ and $k$ PEs will be updated in each record. It will take $O(n'*logn)$ time to find $n'$ time slots by searching \emph{TimeSet}, take $O(n)$ time to find the record for each time slot in the linked list and $O(k)$ time to update $k$ PEs for each record. After updating the records of $t_s$ and $t_e$, it will take $O(1)$ time to remove them if they are redundant. Thus the overall complexity of finding and updating $n'$ records will take $O(n'*(n+k+logn))$ time.

\subsection{Search feasible allocation}
\label{sec:search}
When a new AR request arrives, this operation is performed to check whether there are enough PEs to be allocated for the job. If so, the operation will choose and return the identities of allocated PEs and the start time for the job; Otherwise, it will return null, inferring that there are not enough PEs for the request. This operation is defined as $findAllocation(t_r,t_{du},t_{dl},n_{pe},policy)$ and is shown in Algorithm \ref{alg:findResources}, where $policy$ is the scheduling policy used to choose available PEs and runtime intervals for the request(see Section \ref{sect:strategies}). If \emph{AvailRectList} is empty, the operation will allocate the request to start at $t_r$ and allocate $n$ PEs for it(line \ref{alg3:line:p1}-\ref{alg3:line:p2}); Otherwise, the operation will search for feasible start times(line \ref{alg3:line:p3}), get the maximum availability rectangle of every start time(line \ref{alg3:line:p31}-\ref{alg3:line:p4}) and add them into \emph{availRect}; If finally \emph{availRect} is not empty, inferring there are feasible allocations for the request, the operation will choose an appropriate start time and allocate PEs for the request according to the scheduling policy(line \ref{alg3:line:p5}).

Notably, any time slot in the interval $[t_r, t_{dl}]$ may be an optional start time for the request. This makes it a hard work to check the availability rectangles of resource related to these start times. To simplify this operation and to minimize the possible fragmentation of resources resulting from AR allocations, in the operation of line \ref{alg3:line:p3}, it's suggested to check existing time slots only in the interval $[t_r, t_{dl}]$ and new ones generated by deducting these existing time slots with $t_{du}$. For every optional start time $t_s$, the operation gets free PEs(i.e., $PE_{free}$),  in the interval $[t_s, t_s+t_{du})$. This can be done by iterating through \emph{AvailRectList}. If the number of free PEs is not less than $nPE$, indicating $t_s$ is a feasible start time, and the operation will find the maximum availability rectangle containing $PE_{free}$ and the interval $[t_s, t_s+t_{du})$(line \ref{alg3:line:p40}). Finally, after constructing availability rectangles for all feasible start times, the operation will choose one of them according to $policy$, and return the appropriate start time and $n_{job}$ PEs for the request.

\begin{algorithm}[htbp]
\caption{\em{findAllocation}($t_r$,$t_{du}$,$t_{dl}$,$n_{job}$,$policy$)}
\label{alg:findResources}

\eIf{$AvailRectList$ is empty}
    {
    Let the job to run at $t_r$ and allocate $n$ PEs for it;\;\label{alg3:line:p1}
    return $\{t_r,\textrm{IDs of the $n$ PEs}\}$\;\label{alg3:line:p2}
    }{
    find all feasible start times $\{ST\}$ within $[t_r, t_{dl}-t_{du}]$\;\label{alg3:line:p3}

    \ForEach{element $t_s$ in ${ST}$}
        {\label{alg3:line:p31}

        find the maximum availability resource rectangle $\{T_{begin},T_{end},PE_{free}\}$ of $t_s$\;\label{alg3:line:p40}

        $availRect.add(\{t_s,T_{begin},T_{end},PE_{free}\})$\;}

    \label{alg3:line:p4}


    \uIf{$availRect$ is not empty}
    {
    choose the appropriate $t_s$ and $n_{job}$ PEs according to $policy$ from availability resource rectangles, and return $\{t_s,\textrm{IDs of the $n_{job}$ PEs}\}$\;\label{alg3:line:p5}
    }
    \lElse{return null\;   }\label{alg3:line:p6}
    }

\end{algorithm}

The complexity of \emph{findAllocation()} is as follows. If $AvailRectList$ is empty, the request will be allocated to run at $t_r$ immediately. This will take $O(1)$ time; Otherwise, we can sort the linked list into sorted array list(this will take $O(nlogn)$), and assume there are $p$ feasible start times within $[t_r,t_{dl}]$. For each feasible start time, assume there are $u$ free PEs in its maximum availability rectangle and $v$ neighboring records should be checked to determine the maximum availability rectangle, it will take $O(u*v)$. After getting the maximum availability rectangle of each feasible start time, the information of the rectangles will be used to build a priority queue($O(p)$), in which the selected rectangle according to \emph{policy} will always be in the root($O(1)$). Finally, a group of $n_{job}$ free PEs will be chosen from the selected rectangle
with $u$ free PEs and allocated to the request($O(n_{job}*logu)$). Overall, the complexity of searching and allocating resources for the request will take  $O(p*u*v+nlogn+n_{job}*logu+p)$.

The following example illustrates a typical application of these operations in Figure \ref{fig:fig1}. At $t_0$, there are two running jobs and one reserved job. Assume the running jobs begin to run at $t_0$ and the records in \emph{AvailRectList} are $\{t_0,n1+n2\; PEs\}$,$\{t_1,n1\;PEs \}$,$\{t_3,\o\}$,$\{t_8,n3\;PEs\}$ and $\{t_{10},\o\}$. The following steps illustrate the actions of the above operations.

(1)When a new AR request $\{t_2,t_4-t_2,t_9,n\}$ arrives, the scheduler calls \emph{findAllocation($t_2,t_4-t_2,t_9,n,policy$)} to find available start times and free PEs for the job. Theoretically, it's optional for the AR job to start at any time slots within from the ready time $t_2$ to the latest start time $t_7$. However, we only choose $t_2$,$t_3$,$t_6$ and $t_7$ as feasible start times and neglect any other slots. In this way, we can simplify the searching operation and lower the influence of AR requests on fragmenting resources.

(2)Fortunately, there are enough free PEs for the AR request to begin to run at any of the four start times, and \emph{findAllocation()} will calculate the maximum availability rectangle for every start time and choose one of them for the AR request. For $t_2$,the number of free PEs within the interval $[t_2,t_4)$ are $N-n1$, and the beginning slot and ending slot of the maximum availability rectangle with $N-n1$ free PEs are $t_1$ and $t_8$. For $t_3$, the number of free PEs within the interval $[t_3,t_5)$ are $N$, and the beginning slot and ending slot of the maximum availability rectangle with $N$ free PEs are $t_3$ and $t_8$. In this way,we can get the numbers of free PEs and the beginning slots and ending slots of the maximum availability rectangles of $t_6$ and $t_7$ respectively. Assume \emph{policy} is \emph{PE Worst Fit}(see \emph{Section}\ref{sect:strategies}) and $t_3$ is chosen as the start time and $n$ PEs will be allocated for the request, the operation will return $\{t_3, n \;PEs \}$.

(3)After getting the start time and PEs for the AR request, the scheduler will call \emph{addAllocation($t_3,t_5,n \;PEs$)} to add the reservation into \emph{AvailRectList}. At first,the adding operation updates $\{t_3,\o\}$ to $\{t_3,n \;PEs\}$, and inserts $\{t_5,\o\}$ into \emph{AvailRectList}. Because$\{t_1,n1\;PEs \}$ is the exactly previous record of $\{t_3, n \;PEs\}$ in \emph{AvailRectList} and the $n1$ PEs of $t_1$ are the same as the $n$ PEs of $t_3$, $\{t_3, n \;PEs\}$ will be merged with $\{t_1, n1 \;PEs \}$ and removed from \emph{AvailRectList} .

(4)At $t_1$, job2 finishes, and ${deleteAllocation(t_0,t_1, n2 PEs)}$ will be called to subtract $n2$ PEs from the records within the interval $[t_0,t1)$. The original record of $t_0$ will change from $\{t_0,n1+n2\; PEs\}$ to $\{t_0,n1 PEs\}$, and the original record of $t_1$, i.e., $\{t_1,n1\;PEs \}$, will be merged with the new $\{t_0,n1 PEs\}$ and then be removed. Finally the remaining records in \emph{AvailRectList} are $\{t_1,n1 PEs\}$, $\{t_5,\o\}$, $\{t_8,n3\;PEs\}$ and $\{t_{10},\o\}$.

\section{Scheduling algorithms}
\label{sect:strategies}
If there are more than one allocations that can satisfy the request at the same time, the scheduler will choose one of them based on some criteria. Considering feasible start times themselves and their maximum availability rectangles, we have developed following scheduling strategies to control the allocation of resources for AR requests.

\begin{description}
  \item[First Fit(FF):] the job is allocated to run at the earliest feasible start time.
  \item[PE Best Fit(PE\_B):] the job is allocated to run at the feasible start time with the minimum number of free PEs.

  \item[Duration Best Fit(Du\_B):] the job is allocated to run at the feasible start time, the availability rectangle of which has the minimum duration.

  \item[PE-Duration Best Fit(PEDu\_B):] the job is allocated to run at the start time, the availability rectangle of which has the minimum production of the number of free PEs and the duration.
\end{description}

Different from \textbf{PE\_B},\textbf{DU\_B} or \textbf{PEDU\_B} that tries to choose feasible start time the availability rectangle of which has the minimum number of free PEs, duration or production, we can also construct their corresponding maximum versions, i.e., the \textbf{PE  Worst Fit (PE\_W)} algorithm, the \textbf{Duration Worst Fit (Du\_W)} algorithm and the \textbf{PE-Duration Worst Fit (PEDu\_W)} algorithm.

In practice, it's possible that more than one feasible start times have the same availability rectangle. For example, in Figure \ref{fig:fig1}, $t_3$ and $t_6$ have the same availability rectangle, which has $N$ free PEs within $t_3$ and $t_8$. In such cases, if the maximum availability rectangle was chosen for the request, the earliest feasible start time will be chosen, so as to shorten the waiting time of the job. e.g., in Figure \ref{fig:fig1}, $t_3$, instead of $t_6$, will be chosen by the scheduler when Du\_B or PE\_W is used.

\section{Performance evaluation}

In order to verify the data structure and its operations, and to evaluate the performance of the scheduling strategies, we implemented the data structure and its operations in a discrete event-driven simulator, applied these strategies to schedule AR requests, and analyzed their performance metrics.

The simulator is implemented on the basis of SimJava\cite{simjava}, which is a process based discrete-event simulation package for Java and is originally developed by University of Edinburgh. For its accuracy in simulation, SimJava is widely used to build simulators in many researches. A SimJava simulation is a collection of entities each running in its own thread and they are connected together by ports and can communicate with each other by sending and receiving events. A central system class controls all the entities, advances the simulation time, and delivers the events. In our simulator, we implemented a hierarchal architecture to model cluster or grid-like computing environments and to evaluate the operation and performance of different resource management strategies. The simulator includes entities such as meta-users, meta-schedulers and multiprocessor systems. A meta-user is responsible for generating AR requests and submit them to the job queue of the meta-scheduler, and the meta-scheduler links to multiprocessor entities and manages their availability information via the data structure proposed in this paper and allocate resources according to the scheduling policies. In a multiprocessor entity, a local scheduler entity and multiple processing element entities were created and they are responsible for processing the AR request submitted by the meta-scheduler.

\subsection{Simulation environments}

For experiments based on discrete event-based simulation, a workload is needed to
drive the simulation. However, there are not any workload traces about advance reservation can be
used in this paper directly. In this paper, the LANL-CM5 in Parallel Workload Archive\cite{pwa} and the Feitelson-Lublin model\cite{lublin2003workload} were considered to generate AR jobs with deadline constraints. The LANL-CM5 is a 1024-node Connection Machine CM-5 system and processors are allocated only in powers of 2, with the minimal partition size and the maximal partition size being 32 and 1024 processors respectively. In experiments, the distributions and parameters used in the Feitelson-Lublin model to generate workload were set according to the LANL-CM5 values in \cite{lublin2003workload},following models and parameters were used to control the generating of jobs:

(1)The combined model of arrival process in the Feitelson-Lublin model and its parameters for LANL-CM5 were used to control the arrival of jobs.

(2)The two-stage uniform distribution with parameters ULow,UMed,UHi and Uprob was used to control the sizes of jobs generated. In this distribution,all jobs are parallel,i.e.,the probability for serial jobs are 0, and their sizes are power of 2,with the minimal size of 32(i.e.,ULow=4.5), the maximal size of 1024(i.e.,UHi=10) and $Uprob=0.82$. For the original LANL-CM5 log,UMed is 7. In order to control the sizes of the jobs generated, UMed was set to be 5, 6, 7, 8, and 9 individually in experiments. As UMed changes from 5 to 9, the mean size of jobs increases.

(3)Runtime is an important characteristic of a rigid job. In the Feitelson-Lublin model, the hyper-Gamma distribution was used to model runtimes and a group of parameters were verified to be appealing and representative for each and all workloads. Although the resulting runtimes in this model are discrete, the distribution of which is very different from the distribution of sizes and spans a very large range of values. In the interest of efficient computability and representability,  we made minor modifications for this model to only generate runtime values of 60, 300, 900, 1800, 3600 and 10800. The distribution of these new runtime values were determined by comparing the distribution of estimated runtimes in the original LANL-CM5 records and the distribution of runtimes generated in the model. Moreover, as the size and the runtime of a job are correlated, when UMed changes,the distributions of sizes and runtimes will change. In this way, we can evaluate the performance of different scheduling algorithms as the distributions of sizes and runtimes of jobs change.

By using the the LANL-CM5 workload and the Feitelson-Lublin model, we can generate a series of jobs, each with arrival time, size and duration. In order to add deadline and advance reservation constraints to the resulting jobs, two factors were used:

\begin{itemize}
\item{artime factor($\ge 0$): is used to control the period between the arrival time $t_a$ and the ready time $t_r$ of an AR request. The period is defined as $artime factor*U[0,1]*t_{du}$, where $U[0,1]$ is a random number uniformly distributed in $[0,1]$.}. This parameter is set based on \cite{heine2005}.
\item{deadline factor($\ge 0$): is used to control the job's deadline, which is defined as $t_r+(1+deadline factor*U[0,1])*t_{du}$. If this factor is zero, the deadline is immediate, i.e., $t_{dl} = t_r+t_{du}$; Otherwise, the deadline is general, i.e., $t_{dl} \ge t_r+t_{du}$.}
\end{itemize}

With these parameters, we can generate jobs with deadlines and advance reservation requests from a workload trace. For AR jobs, as the values of \emph{artime factor} and \emph{deadline factor} increase, the flexibility of scheduling will increase, and the resource competition between AR jobs will be alleviated. Based on the influence of these parameters on AR jobs, these factors were combined together as \emph{\{artime factor,deadline factor\}}, and five pairs of values, i.e., $\{1,1\}$,$\{2,2\}$,$\{3,3\}$,$\{4,4\}$ and $\{5,5\}$, were used to generate low-, middle- and high-flexibility AR jobs.

In order to generate workloads with different distribution of inter-arrival times and further to investigate the performance of strategies under different system load, \emph{arrival factor(af in short)} is defined and used as follows: for a job in a given workload with arrival time $t_{s0}$, its new arrival time will be $t_{s0}/arrival factor$. In this way, we can control the arrival of jobs and thus control system load. In experiments, $af=1$ is set as default.

In experiments, following two metrics were used to evaluate the performance of different scheduling strategies:

(1)\emph{Acceptance rate}: is the percentage of reservations that are accepted because their requirements can be satisfied, which indicates the ability to accommodate AR request.

(2)\emph{Average slowdown}: the slowdown of an AR job is the response time of the job normalized by the running time, i.e., $(waiting\; time +runtime)/runtime$, where waiting time is the difference between the ready time and the actual start time. This measures how much slower the job ran due to conflicts with other competing jobs and it seems more reasonable than the waiting time to capture the user's expectation that a job's waiting time will be proportional to its runtime. Average slowdown is the average value of slowdowns of all accepted AR jobs, which indicate how well the scheduling algorithm can satisfy the user's expectations on the execution of the job.

\subsection{Experimental results}
\label{results}
In experiments, we investigated the performance of the scheduling strategies against different job sizes and durations, different arrival factors and different \emph{\{artime factor,deadline factor\}} values.For each experiment, $10^4$ jobs were submitted to the scheduler for the results, and we have obtained 95\% confidence intervals for them.

\subsubsection{Results for different job sizes and durations}

\begin{figure}[htp]
\centering
\includegraphics[width=0.75\textwidth]{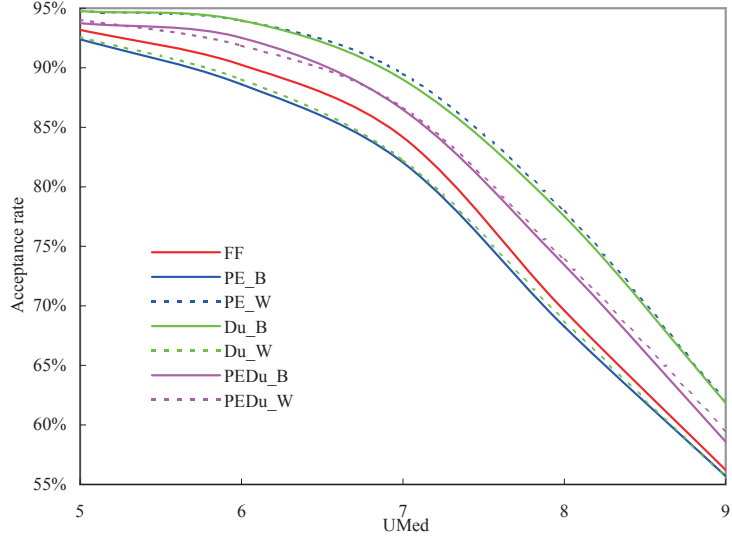}
\caption{Acceptance rate vs job size control parameter UMed}
\label{fig:fig41}
\end{figure}

\begin{figure}[htp]
\centering
\includegraphics[width=0.75\textwidth]{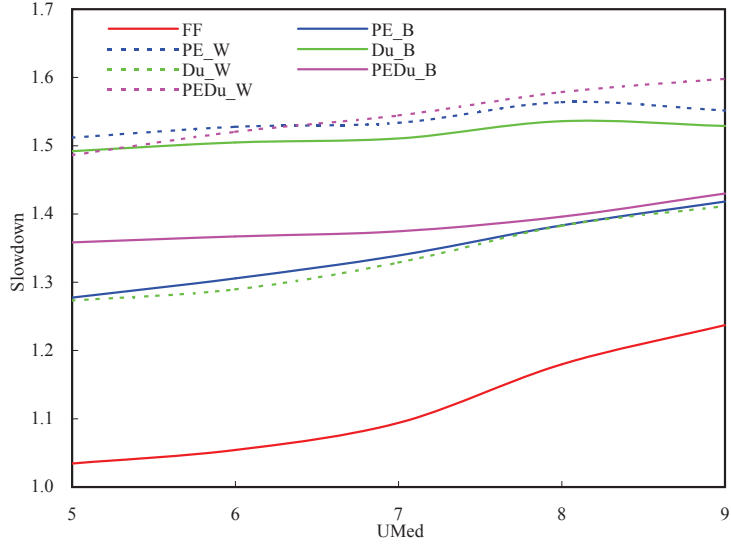}
\caption{Slowdown vs job size control parameter UMed}
\label{fig:fig42}
\end{figure}

Figure \ref{fig:fig41} and Figure \ref{fig:fig42} present the acceptance rate and the average slowdown of different scheduling strategies for workloads with different UMed values respectively. In experiments, \emph{arrival factor} is 1 and  \emph{\{artime factor,deadline factor\}} is $\{3,3\}$ as default. As shown in Figure \ref{fig:fig41}, when UMed changes from low to middle and high, the acceptance rates of all strategies decrease gradually. This result is in agreement with intuition: as UMed increases, the mean size and the mean duration of jobs increase, thus demanding more resources to accommodate them and intensifying competition of available resources between jobs.

In Figure \ref{fig:fig41}, there are three groups of algorithm with almost identical behavior: $PE\_W$ and $Du\_B$, $PEDu\_B$ and $PEDu\_W$, and $PE\_B$ and $Du\_W$. Among them, except the $PE\_B$ algorithm and the $Du\_W$ algorithm, all other four strategies outperform \emph{FF}. Moreover, the $PE\_W$ algorithm and the $Du\_B$ algorithm in the first group perform much better than \emph{FF} and clearly become the best strategies for all UMed values. Notably, except for the almost identical behavior of $PEDu\_B$ and $PEDu\_W$, the performance of $PE\_B$ and $PE\_W$, and $Du\_B$ and $Du\_W$ are quite different. Based on this results, we cannot draw a deterministic conclusion that PE-based strategies are better or worse than duration-based ones, or best fit-based strategies are always better or worse than equivalent worst fit-based ones. This can be explained by the fact that, for an idle period of resource,the influences of the number of its PEs and its duration on accommodating new jobs are different.

Now turn to Figure \ref{fig:fig42}, which plots the average slowdown of different strategies for workload with different UMed values. When UMed changes from low to middle and high, the average slowdown of all strategies  increase in general. This can be explained by the fact: as UMed increases, the mean size and the mean duration of jobs increase. For a new job,no matter under which algorithm, it will experience a longer waiting time before execution. As we can see, the jobs with $FF$ experience the lowest average slowdown. This can be easily explained that $FF$ always chooses the earliest feasible period and thus minimizing the waiting time of jobs. For the other strategies, the performance of $PE\_W$ and $Du\_B$, and $PE\_B$ and $Du\_W$ are similar again as in Figure \ref{fig:fig41}. However, the performance of $PEDu\_B$ and $PEDu\_W$ are surprisingly different.

\subsubsection{Results for different system load}
In both \cite{lublin2003workload} and the aforementioned experiments,UMed is typically set to 7.
In following experiments, we will investigate the performance of the strategies against different arrival factor values with UMed=7 and  \emph{\{artime factor,deadline factor\}}=$\{3,3\}$.

\begin{figure}[htp]
\centering
\includegraphics[width=0.75\textwidth]{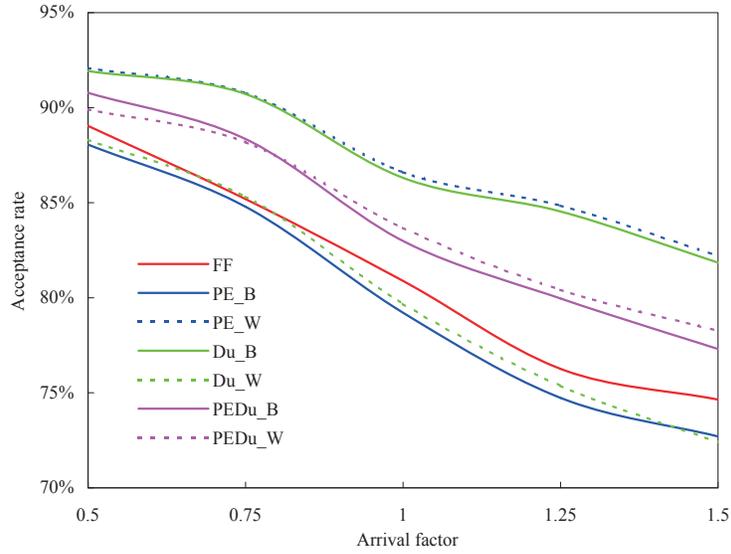}
\caption{Acceptance rate vs arrival factor with UMed=7}
\label{fig:accept2}
\end{figure}

\begin{figure}[htp]
\centering
\includegraphics[width=0.75\textwidth]{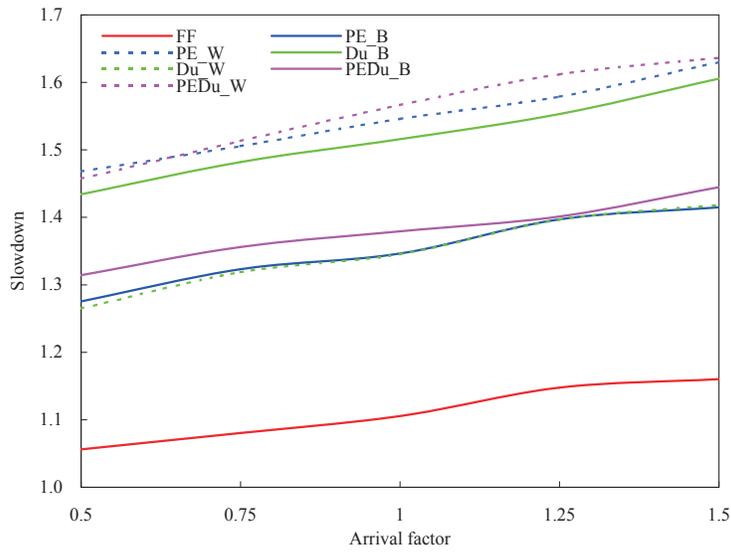}
\caption{Slowdown vs arrival factor with UMed=7}
\label{fig:slowdown2}
\end{figure}

Figure \ref{fig:accept2} and \ref{fig:slowdown2} illustrate the acceptance rate and the average slowdown of the strategies as arrival factor changes from 0.5 to 1.5, in step of 0.25.As expected, as the value of arrival factor increases, acceptance rates and slowdowns of all strategies degrade in both Figures.This agrees with the fact that as the value of \emph{arrival factor} increases, the number of AR requests submitted within a given period will increase,thus the competition of resources among jobs will intensify,and the acceptance rate will decrease. For accepted AR requests, they also tend to experience longer waiting time,for there will be more jobs allocated in their expected execution periods as the value of \emph{arrival factor} increases.

By comparing the results in Figure \ref{fig:fig41} and \ref{fig:accept2} and the results in Figure \ref{fig:fig42} and \ref{fig:slowdown2} respectively,it can be seen clearly that the relative performance of the scheduling algorithms in both experiments are similar. Based on the results in Figure \ref{fig:fig41}-\ref{fig:slowdown2}, we can conclude that job sizes and durations and system load will impact the performance metrics of all the scheduling algorithms and the performance perceived by the users clearly: as job sizes and durations and system load increase, the acceptance rate and the average slowdown for all algorithms will degrade, and AR jobs will experience lower acceptance rate and higher waiting time.

\subsubsection{Results for different scheduling flexibilities}

\begin{figure}[htp]
\centering
\includegraphics[width=0.75\textwidth]{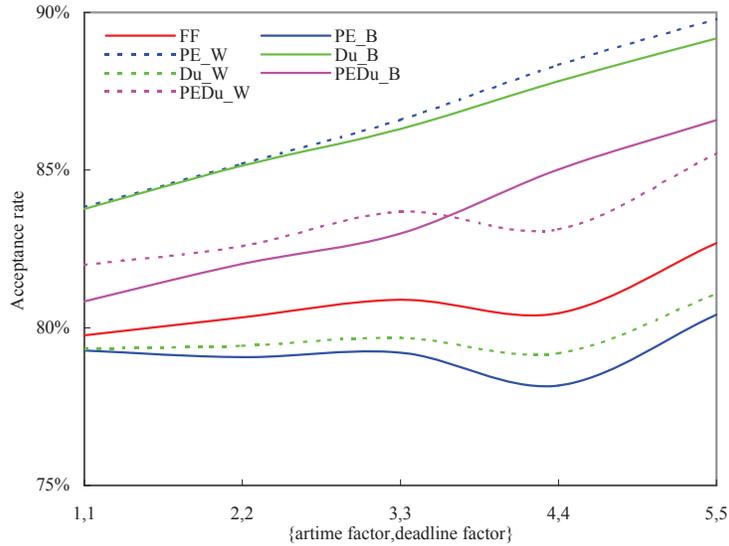}
\caption{Acceptance rate vs $\{artime factor, deadline factor\}$ with UMed=7}
\label{fig:accept3}
\end{figure}

\begin{figure}[htp]
\centering
\includegraphics[width=0.75\textwidth]{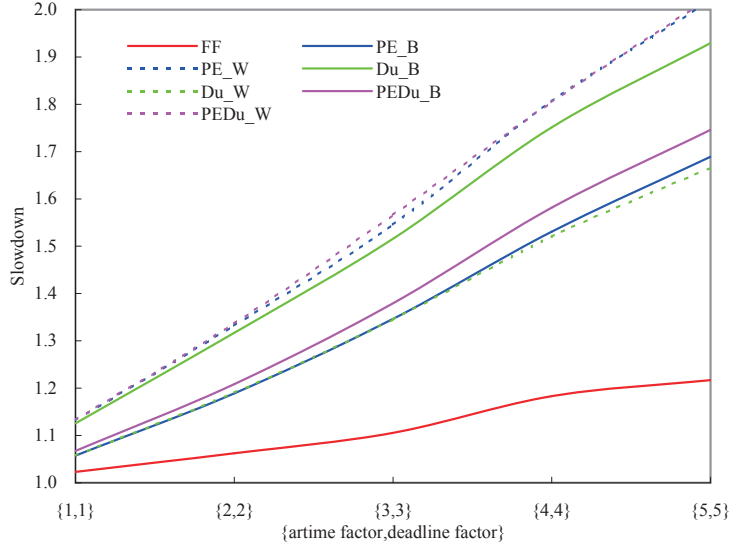}
\caption{Slowdown vs $\{artime factor, deadline factor\}$ with UMed=7}
\label{fig:slowdown3}
\end{figure}

Figure \ref{fig:accept3} and \ref{fig:slowdown3} present the acceptance rate and  average slowdown of different scheduling algorithm when the values of \emph{\{artime factor,deadline factor\}} change.
As shown in Figure \ref{fig:accept3}, when the values of \emph{\{artime factor,deadline factor\}} change from low to middle and high, the acceptance rates of $PE\_W$, $Du\_B$ and $PEDu\_B$ increase almost linearly. This behavior indicates that their acceptance ability are stable throughout the range of flexibilities considered in this study.However, the performance improvements of other four strategies are not stable, especially at $\{4,4\}$, indicating that they are sensitive to the degree of scheduling flexibility.  Among all strategies, $PE\_W$ become the best algorithm again with the highest acceptance rate. It presents better performance than $Du\_B$ as the flexibility of scheduling increases and defeats other strategies easily throughout the range of values.

Figure \ref{fig:slowdown3} presents the average slowdowns of strategies with different scheduling flexibilities. As the scheduling flexibility increase, the average slowdowns of all strategies increase, which agrees with the intuition that the more flexibility an AR request has in scheduling, the longer the waiting time and larger slowdown will be.  Moreover, the relative performance of the curves are similar to the others observed earlier: $FF$ is always the one with smallest values of slowdown by allocating AR jobs to run as early as possible.

\section{Conclusions and discussions}
\label{conclusion}

In this paper,we discuss about the scheduling model and algorithms for parallel AR jobs with deadline. We proposed a new data structure and a set of operations to organize the availability of multiprocessor systems for single- and/or multiple-processor advance reservation requests with immediate or general deadline constrains in a way that enables efficient search and update operations, formulated a suite of scheduling policies for the data structure to allocate resources for AR requests, and investigated their performance via simulation.
Based on a comprehensive performance evaluation study of the scheduling policies with simulation, it's shown that job sizes and durations, system load and the flexibility of scheduling will impact the performance metrics of all the scheduling algorithms. Among them, the $PE\; Worst\; Fit$ algorithm becomes the best algorithm for the scheduler with the highest acceptance rate of AR requests, and the jobs with the $First\; Fit$ algorithm experience the lowest average slowdown. The simulator and the simulations verified that the data structure, its operations and the scheduling policies are efficient and effective in such computing environments, and can be used in practice. Moreover, because the data structure can support different scheduling policies in a flexible way, other scheduling policies can be easily integrated in the system.

In the research of the data structure,its operations and the scheduling policies, we assume that the resources are homogeneous and the jobs are rigid. However, They can be extended to support heterogeneous resources and malleable jobs in the future. If the system is heterogeneous,i.e., the capacities of the resources in the system are not the same, we can standardize the capacities of the resources and the requirements of the jobs by using a 'standard' resource. In this way, the capacity of each resource and the requirement of each job are described by referring to the standard resource, and we can organize the 'standardized' capacities of the resources in the data structure, and allocate 'standardized resources' for the jobs with 'standardized' requirements. On the other hand, for malleable jobs, their requirements on the number of PEs and durations are not fixed. If a malleable job's requirement of the number of PEs changes, its duration will change along. However, in the $findAllocation(t_r,t_{du},t_{dl},n_{job},policy)$ algorithm in SubSection \ref{sec:search}, the number of PEs(i.e., $n_{job}$) and the time-related constraints(i.e., $t_r$,$t_du$,$t_dl$) are rigid. To support malleable jobs, the malleable requirements on the number of PEs and time-related parameters of a job should be 'translated' into a group of rigid ones, then those rigid parameters can be used to find resources for the jobs by using  the $findAllocation(t_r,t_{du},t_{dl},n_{job},policy)$ algorithm. Additionally, some new criteria should be designed to choose an allocation for the malleable job among the group of rigid parameters. How to 'translate' the requirements of a malleable AR job with deadline constraint into a group of rigid parameters is also a problem to be considered. In the future, we plan to investigate the problems for heterogeneous resources and malleable jobs in more detail.

\section*{Acknowledgements}
This research is supported in part by the Natural Science Foundation of China under grant number 60663009, the Training Programme Foundation for Young Key Teachers of Yunnan University and the Research Foundation of Yunnan University under grant number 2009F30Q. Also we would like to thank the reviewers for their valuable suggestions and comments on this paper.

\bibliographystyle{elsarticle-num}

\end{document}